\documentclass[10pt,conference]{IEEEtran}

\usepackage{float}
\usepackage{graphicx}
\usepackage{clrscode}
\usepackage{paralist}
\usepackage{tikz}
\usepackage{soul} 
\usepackage{color}
\usepackage{cite}
\usepackage{multicol}
\usepackage{comment}
\usepackage{amsmath, amssymb, mathrsfs, amsfonts}
\usepackage{amsthm}
\usepackage{mathtools}
\usepackage{epsfig, epstopdf}
\usepackage{subfig}
\usepackage{algorithm}
\usepackage{enumitem}
\usepackage{algorithmic}

\usepackage{lipsum}
\usepackage{pbox}
\usepackage{array}
\usepackage{tablefootnote}
\usepackage{caption}
\usepackage[square,numbers,sort&compress]{natbib}

\usepackage{multicol}
\usepackage{dblfloatfix}

\usetikzlibrary{arrows,automata}

\allowdisplaybreaks

\newtheorem{lemma}{Lemma}

\newcommand{\RNum}[1]{\uppercase\expandafter{\romannumeral #1\relax}}

\begin{document}

\title{Outage Tradeoff Analysis in a Downlink Integrated
Sensing and Communication Network}

\author{
		\IEEEauthorblockN{Marziyeh Soltani, Mahtab Mirmohseni, and Rahim Tafazolli \\
		\vspace*{0.5em}
			\IEEEauthorblockA{
                              5GIC and 6GIC, University of Surrey, United Kingdom\\
                              }}}

\maketitle
\begin{abstract}
This paper aims to analyze the stochastic performance of a multiple input multiple output (MIMO) integrated
sensing and communication (ISAC) system in a downlink scenario, where a base station (BS) transmits a dual-functional radar-communication (DFRC) signal matrix, serving the purpose of transmitting communication data to the user while simultaneously sensing the angular location of a target. The channel between the BS and the user is modeled as a random channel with Rayleigh fading distribution, and the azimuth angle of the target is assumed to follow a uniform distribution. We use a maximum ratio transmission (MRT) beamformer to share resource between sensing and communication (S \& C) and observe the trade-off between them. We derive the approximate probability density function (PDF) of the signal-to-noise ratio (SNR) for both the user and the target. Subsequently, leveraging the obtained PDF, we derive the expressions for the user's rate outage probability (OP), as well as the OP for the Cramer-Rao lower bound (CRLB) of the angle of arrival (AOA). In our numerical results, we demonstrate the trade-off between S \& C, confirmed with simulations.
\end{abstract}
\section{Introduction}\label{introduction}
Recently, there has been a noticeable shift in the approach to radar sensing and wireless communication systems, referred to as integrated sensing and communications (ISAC), which is expected to play a pivotal role in advancing next-generation wireless networks \cite{SeventyYearsofRadarandCommunications}. ISAC is required in various emerging applications, including vehicular communication, internet of things (IoT) applications like smart city infrastructure (such as traffic and speed monitoring, video surveillance), and smart industry \cite{Integratedtoward}. By leveraging shared resources such as wireless spectrum, hardware platforms, and energy consumption, ISAC offers significant advantages for both sensing and communication (S \& C). Due to this resource sharing, there is an inherent tradeoff in ISAC, and in order to design efficient ISAC system, the fundamental
communication-sensing performance tradeoff should be fully
understood \cite{ASurveyonFundamentalLimits}.

The recent comprehensive review paper \cite{ASurveyonFundamentalLimits} and its referenced literature primarily analyzed the performance of ISAC from an information-theoretic perspective, overlooking the impact of channel fading and its statistical properties. However, more recent studies have delved into the probabilistic behavior of ISAC. These studies can be classified into two categories. In the first category, single transmission antenna is utilized for the base station (BS). These works predominantly consider the detection probability as the sensing performance metric \cite{OnthePerformanceofUplinkandDownlink, PerformanceAnalysisandPowerAllocationforCooperative, WaveformDesignandPerformanceAnalysisforFullDuplex, PerformanceAnalysisandPowerAllocation, FundamentalDetectionProbabilityvsAchievableRate}. In contrast \cite{NOMAISACPerformanceAnalysisandRateRegion} utilizes the sensing rate, defined as the sensing mutual information per unit time, and \cite{RethinkingthePerformanceofISACSystem} employs the ergodic range Cramer-Rao lower bound (CRLB) as the sensing performance metric. In the second category, multiple transmission antennas are employed at the BS. In \cite{MIMOISACPerformanceAnalysis}, the authors present precoding designs for three scenarios: sensing-centric design, communications-centric design, and Pareto optimal design. For each scenario, they derive diversity orders and high signal-to-noise ratio (SNR) slopes (i.e., asymptotic behaviour) for both sensing rate and communication rate. \cite{Aunifiedperformanceframeworkfor} proposes a unified performance framework for ISAC systems, by utilizing the Kullback-Leibler divergence (KLD). This framework enables a unified evaluation of the error rate performance of users and the detection performance of targets. In \cite{PerformanceofDownlinkandUplinkIntegratedSensing}, the authors analyze the communication rate by considering diversity orders and high SNR slopes, without considering any beamforming vector. Furthermore, they unveil the high signal-to-noise ratio slopes for both the communication rate and sensing rate. In \cite{PerformanceAnalysioftheFullDuplexJoint}, the authors focuses on designing the beamforming vector based solely on the user channel, while neglecting the target channel. The authors derive the rate of the user and the estimation rate, analogous to the communication rate for sensing operation, as performance metrics for both communication and sensing.

\textbf{Our Contribution:} 
This paper aims to evaluate the fundamental stochastic performance limits of a downlink multiple input multiple output (MIMO) ISAC system. We focus on the trade-off between S \& C, by considering the influence of channel randomness between the user and the BS, as well as between the target and the BS. The angle of the target is assumed to follow a uniform distribution, while the user channel is modeled as a circularly-symmetric complex Gaussian random vector. Two key metrics are considered to reveal the S \& C trade-off: the outage probability (OP) of the user for communication rate, and the OP of the target, denoted as $P(\text{CRB}(\theta) > \epsilon)$, for sensing accuracy. To enhance both communication rates and radar estimation accuracy, a dual functional precoding matrix is employed instead of solely emphasizing communication or sensing, as discussed in \cite{CrameRaoBoundOptimizationforJoint}. Instead of optimizing the beamforming vector for data precoding at the BS, our approach assumes a predefined beamforming vector with constant parameters and derives the fundamental limits of above metrics. Then, through simulations, the impact of the parameters on performance metrics is examined. These parameters show the compromise between S \& C through resource  sharing. We calculate the OP of the user and OP of the target by first deriving the signal-to-interference-plus-noise ratio (SINR) and CRLB expressions. The stochastic behavior of SINR and CRLB are due to some joint random variables (RVs) (which depend on the channel and the direction of departure (DoD)). We derive the probability density function (PDF) for these RVs, which enables us to characterize the performance metric probabilities on SINR and CRLB, since they are functions of these RVs. The numerical and simulation results reveal a fundamental trade-off between S \& C. In fact, allocating more power towards the user direction leads to degradation in the user's OP, indicating an increase in the user rate. However, it also results in a increase in the OP the target, implying higher error in the angle estimation. Furthermore, the simulation demonstrates that the system's performance in both S \& C aspects improves with an increase in BS power.

\textbf{Differences between this paper and recent works:} In contrast to previous works where the BS only has a single transmission antenna, \cite{OnthePerformanceofUplinkandDownlink}-\cite{RethinkingthePerformanceofISACSystem}, this paper considers a scenario with multiple transmitting and receiving antennas at the BS. This enables us to precode data at the BS before transmission in a desired direction which is essential to observe the trade-off between S \& C. This capability is not possible in the single antenna scenario. However, assuming a beamforming vector in the direction of both the target and the user poses a great challenge in deriving the performance metrics due to the correlation that arises between the user channel, target channel, and the beamforming vector.
In comparison to works that involve multiple antennas at the BS, either the beamforming vector is not considered \cite{PerformanceofDownlinkandUplinkIntegratedSensing}, or it is based solely on the user channel \cite{PerformanceAnalysioftheFullDuplexJoint} (sending a separated downlink signal and radar probing signals without considering a dual functional precoding matrix), or it is designed based on the assumption of knowing the eigenvalues of the covariance matrix of each column of the target response matrix \cite{MIMOISACPerformanceAnalysis}, without considering the angle of the target. Furthermore, none of these works have considered the OP of the target as a sensing performance metric.

\textbf{Notation:} We use letters, boldface lowercase, and boldface uppercase to denote scalar quantities, vectors and matrices, respectively. $P(.), f_x(.),$ $ E[.]$ represent the probability, the PDF, and the expectation, respectively. $\mathbf{X}^{M \times N}$, $\mathbf{X}^T, \mathbf{X}^H$ and $\mathbf{X}^*$ are matrix with $M$ rows and $N$ columns, the transpose, the Hermitian transpose and the conjugate of $\mathbf{X}$, respectively. The Euclidean norm of a vector is $\parallel. \parallel$ and the norm of a complex number is $|.|$. $\mathcal{CN}(.,.)$ denotes a circularly-symmetric complex Gaussian distribution, and $\mathcal{N}_3(\mathbf{\mu},\mathbf{\Sigma})$ is a trivariate normal distribution with mean $\mathbf{\mu}$ and covariance matrix $\mathbf{\Sigma}$. $\mathbb{C}$ and $\mathbb{R}$ are sets of complex and real numbers. $\overset{d}\rightarrow$ shows convergence in distribution, and \text{tr($A$)} is trace of matrix $A$.
\section{System Model}\label{systemmodel}
We assume a BS with $N$ transmitting
antennas and $M$ receiving antennas which aims to serve a communication single-antenna user in the downlink while sensing a target located at a point far away from the BS (point target). We consider a
mono static radar setting: the direction of arrival
(DoA) and the DoD are the same. The BS sends $\mathbf{X} \in \mathbb{C}^{N\times L}$, a narrow-band dual functional radar communication signal matrix, defined by:
\begin{equation}
\mathbf{X}=\sqrt{p_t}\mathbf{w}\mathbf{s}\label{x},
\end{equation}
where $L>N$ is the length of the radar pulse/communication
frame, $\mathbf{w} \in \mathbb{C}^{N \times 1}$
is the beamforming vector, and $\mathbf{s} \in \mathbb{C}^{1 \times L}$ is the white Gaussian
signaling data stream
 for the user with unit power, $\frac{1}{L}\mathbf{s}\mathbf{s}^H\approx1$ when $L$ is sufficiently large (\cite{CrameRaoBoundOptimizationforJoint}). $p_t$ is the transmit power of the BS. We assume that $\mathbf{w}$ lies in the complex span of the channel vector of the users and target. Thus, we have:
\begin{equation}
\mathbf{w} =\frac{b_1\mathbf{h}+b_2\mathbf{a}}{\parallel b_1\mathbf{h}+b_2\mathbf{a} \parallel}\label{w},
\end{equation}
where $\mathbf{h}=[h_1\quad h_2 ...h_N]^T \in \mathbb{C}^{N \times 1}$, is the channel vector between the BS and the user, whose elements are independent and identically
distributed (i.i.d.) with distribution $\mathcal{CN} (0,1)$. More precisely, element $i$-th can be expressed as $h_i=m_i+jn_i$ with $m_i$, $n_i \sim \mathcal{N}(0,\frac{1}{2})$. For all $i$, the elements $m_i$ and $n_i$ are i.i.d. Also, by assuming even number of transmitting antennas, the transmit array steering vector is: $\mathbf{a}(\theta) = [e^{-j\pi \sin(\theta)\frac{N-1}{2}} ... e^{j\pi \sin(\theta)\frac{N-1}{2}}] \in \mathbb{C} ^{N \times 1}$, where $\theta$ is the azimuth angle of the target relative to BS, which is uniformly distributed in the interval $[0,\pi]$. $b_1,b_2 \in \mathbb{C}$ are two constant numbers which can be optimized to maximize the SINR of the user or minimize the CRLB. We express $i$-th element of $\mathbf{a}(\theta)$ as $a_i=e^{-jf_i}$, where $f_i=\pi \sin(\theta)\frac{N-(2i-1)}{2}$.
We note that if we had only user or target, maximum ratio transmission (MRT) beamformer, which is $\frac{\mathbf{h}}{||\mathbf{h}||}$ for the user and $\frac{\mathbf{a}}{||\mathbf{a}||}$ for the target, would be optimal, which means the SINR of the revived signal would be maximum. We note that although instantaneous $\mathbf{h}$ and $\mathbf{a}(\theta)$ are known at the BS, outage probability is a suitable performance metric, as the channel vectors are random.
The received signal at the user is,
\begin{equation}
\mathbf{y}_u=\mathbf{h}^H \mathbf{X}+\mathbf{z}_u, \label{yu}
\end{equation}
$\mathbf{z}_u\in \mathbb{C}^{1 \times L}$ is the additive white Gaussian noise
(AWGN) vector where each of its elements has the distribution of the form $\mathcal{CN} (0,\sigma^2_u)$. When the BS transmits $\mathbf{X}$ to sense the target, it receives back
the reflected echo signal matrix as,
\begin{equation}
\mathbf{Y}_r=\alpha \mathbf{b}(\theta)\mathbf{a}(\theta)^H \mathbf{X}+\mathbf{Z}_r,\label{yr}
\end{equation}
where $\mathbf{Z}_r\in \mathbb{C}^{M \times L}$ is AWGN matrix which its elements being i.i.d. and having the distribution of the form $\mathcal{CN} (0,\sigma^2_r)$. $b(\theta) \in \mathbb{C}^{M \times 1}$ is the receive array steering vector and $\alpha \in \mathbb{C}$ represents the reflection coefficient.
Based on (\ref{yu}) and (\ref{x}), the user's SINR is,
\begin{align}
\text{SINR}=
\frac{p_t}{\sigma^2_u} |\mathbf{h}^H \mathbf{w}|^2\label{sinr}.
\end{align}
\begin{figure*}[t]
\normalsize
\begin{align}
\text{CRB}(\theta)=\frac{\sigma^2_R \text{tr}(\mathbf{A}^H(\theta) \mathbf{A}(\theta) \mathbf{R}_x)}{2 \mid \alpha \mid ^2 L (\text{tr}(\mathbf{A}^H(\theta) \mathbf{A}(\theta) \mathbf{R}_x) \text{tr}(\mathbf{A}^{.H}(\theta) \mathbf{A}^.(\theta) \mathbf{R}_x)-\mid \text{tr}(\mathbf{A}^{.H}(\theta) \mathbf{A}(\theta) \mathbf{R}_x)\mid^2)} \label{crb}
\end{align}
\hrulefill
\end{figure*}
In Appendix C of \cite{TargetDetectionandLocalization}, the CRLB for a given $\theta$ was derived as (\ref{crb}),
where $\mathbf{A}(\theta)= \mathbf{b}(\theta) \mathbf{a}^H(\theta)$; $ \mathbf{R}_x=\frac{1}{L}\mathbf {X}\mathbf{X}^H\approx p_t\mathbf{w}\mathbf{w}^H$ is the sample covariance matrix of $\mathbf{X}$. Inserting (\ref{x})- (\ref{sinr}) into (\ref{crb}) and after some algebraic manipulation, the \text{CRB}($\theta$) is simplified as,
\begin{align}
\text{CRB}(\theta)=\frac{\sigma_r^2}{2 L p_t|\alpha|^2 || \mathbf{b}^. ||^2 | \mathbf{a}^H \mathbf{w} | ^2}.\label{cramer}
\end{align}
\section{Performance Analysis}
In this section, we analyse the system performance and calculate the OP the user, i.e., $P(\text{SINR}<\gamma)$, and OP of the target, i.e., $P(\text{CRB}>\epsilon)$.
\subsection{OP of the user}\label{opofuser}
Based on (\ref{sinr}) and (\ref{w}), the SINR of the user is:
\begin{align}
\text{SINR}&=\frac{p_t}{\sigma^2_u}\frac{|\mathbf{h}^H (b_1\mathbf{h}+b_2\mathbf{a})|^2}{|b_1\mathbf{h}+b_2\mathbf{a}|^2}\nonumber\\
&=\frac{p_t}{\sigma^2_u}\frac{|\sum_{i=1}^{N}(b_1|h_i|^2+b_2h^*_ie^{-jf_i})|^2}{\sum_{i=1}^{N}|b_1h_i+b_2e^{-jf_i}|^2)}\nonumber\\
&\overset{(a)}{=}\frac{p_t}{\sigma^2_u}\frac{(\sum_{i=1}^{N}x_i)^2+(\sum_{i=1}^{N}y_i)^2}{(\sum_{i=1}^{N}k_i)}\nonumber\\&\overset{(b)}{=}\frac{p_t}{\sigma^2_u}\frac{X^2+Y^2}{K},\label{sinr2}
\end{align}
where ($a$) is due to defining random variables $x_i=\mathcal{R}(b_1|h_i|^2+b_2h^*_ie^{-jf_i})$, $y_i=\mathcal{I}(b_1|h_i|^2+b_2h^*_ie^{-jf_i})$, and $k_i=|b_1h_i+b_2e^{-jf_i}|^2$, in which $\mathcal{R}$ and $\mathcal{I}$ indicates real and imaginary parts; ($b$) is due to defining $X=\sum_{i=1}^{N}x_i$, $Y=\sum_{i=1}^{N}y_i$ and $K=\sum_{i=1}^{N}k_i$. Therefore, the OP of the user, $P_u$, is:
\begin{align}
P_u&=P(\text{SINR}<\gamma)=P(\frac{p_t}{\sigma^2_u}\frac{X^2+Y^2}{K}<\gamma)\nonumber\\&\overset{(a)}{=}E_{\theta}\{P(\frac{p_t}{\sigma^2_u}\frac{X^2+Y^2}{K}<\gamma)|\theta\}\nonumber\\&=\int_{0}^{\pi}P(\frac{p_t}{\sigma^2_u}\frac{X^2+Y^2}{K}<\gamma)|\theta)f_{\theta}(\theta)d\theta,\label{outage}
\end{align}
, where ($a$) follows by conditioning on $\theta$. Thus, in order to calculate the inner probability, we need to derive the joint PDF of $X$, $Y$, and $K$. We note that $X$, $Y$, and $K$ (also $x_i$, $y_i$, and $k_i$) are not independent, as they are functions of $h_i$ and $f_i$. By conditioning on $\theta$, $f_i$, $\forall i=1,...,N$, will be treated as constant in the following. We define $N$ random vectors, $\mathbf{d}_i=[x_i, y_i, k_i]^T \in \mathbb{R}^{3 \times 1}$, $\forall i=1,...,N$. For any pair of $j$ and $i\neq j$, the vectors $\mathbf{d}_j$ and $\mathbf{d}_i$ are independent from each other because $h_i$s are i.i.d.; more precisely, these vectors are i.i.d. Thus, by using multidimensional central limit theorem (CLT) \cite{enwiki:1155685628}, when $N$ is large \footnote{Section \ref{simulations} shows that for $N>9$, multidimensional CLT holds. Moreover, in \cite{Aunifiedperformanceframeworkfor} and with the help of simulation, the authors show that the accuracy of CLT for a one dimensional random variable holds for $N>8$.} (which holds for the case of MIMO ISAC, due to using large antenna arrays), we have:
\begin{align}
\sqrt{N}[\frac{1}{N}(\sum_{i=1}^{N}\mathbf{d}_i)-\mathbf{\mu_d}]\overset{d}{\rightarrow} \mathcal{N}_3(\mathbf{0},\mathbf{\Sigma_d}),
\end{align}
which means $\sum_{i=1}^{N}\mathbf{d}_i\overset{d} \rightarrow \mathcal{N}_3(N\mathbf{\mu_d},N\mathbf{\Sigma_d})$, 
where $\mathbf{\mu_d}$ and $\mathbf{\Sigma_d}$ are mean vector and covariance matrix of $\mathbf{d}_i$ (the same for all $i=1,...,N$). Therefore, $[X, Y, K]^T\overset{(d)}{\rightarrow} \mathcal{N}_3(N\mathbf{\mu_d},N\mathbf{\Sigma_d})$. Thus, by finding $\mathbf{\mu_d}$ and $\mathbf{\Sigma_d}$ with the aid of Lemma \ref{lemma1} (proof in Appendix \ref{lemma1p}), the joint PDF of $X$, $Y$, and $K$ is derived.
\begin{lemma}\label{lemma1}
By defining $b_1=|b_1|e^{j\phi_1}$ and $b_2=|b_2|e^{j\phi_2}$, $\mathbf{\mu_d}$ and $\mathbf{\Sigma_d}$ are calculated as (\ref{mud}) and (\ref{sigmad}), respectively:
\begin{align}
&\mathbf{\mu_d}=[|b_1|\cos(\phi_1), |b_1|\sin(\phi_1), |b_1|^2+|b_2|^2]^T.\label{mud}
\end{align}
\end{lemma}
We note that as derived in Lemma \ref{lemma1}, the joint conditional PDF of $X$, $Y$, and $K$ is independent of $\theta$ and $\phi_2$. Using this and the assumption of uniform distribution for $\theta$, (\ref{outage}) will be:
\begin{align}
P_u&=\!\!P(\frac{p_t}{\sigma^2_u}\frac{X^2+Y^2}{K}\!\!<\!\!\gamma)\!\!=\!\!\!\iiint_{\!\!\frac{p_t}{\sigma^2_u}\frac{X^2+Y^2}{K}<\gamma}\!\!\!\!\!\!\!\!\!\!\!f(X,Y,K)\,dX\,dY\,dK,\label{outage2}
\end{align}
where $f(X,Y,K)$ is the PDF of a trivariate normal distribution with a mean vector of $N\mathbf{\mu_d}$ and a covariance matrix of $N\mathbf{\Sigma_d}$ \footnote{We remark that by deriving $f(X,Y,K)$, we can also calculate the ergodic rate as $E[\log^{1+\text{SINR}}]=\int_{0}^{\infty}P(\frac{X^2+Y^2}{K}>\frac{\sigma_t(2^t-1)}{p_t})dt=\int_{0}^{\infty}\iiint_{\frac{X^2+Y^2}{K}>\frac{\sigma_t(2^t-1)}{p_t}}f(X,Y,K)\,dX\,dY\,dKdt$.}. Integrating a general multivariate normal PDF over arbitrary interval has no general analytical
expression, and we must use numerical methods such as the numerical method of ray-tracing \cite{Amethodtointegrate}.
However, since the domain is quadratic, we simplify the quadratic form into a weighted sum of non-central chi-square variables. Then, we calculate this integral \cite{ProbabilityContentofRegions, Amethodtointegrate}. First, we write the domain $\frac{p_t}{(\sigma^2_u)}\frac{X^2+Y^2}{K}<\gamma$ in a quadratic form. Since all the constant parameters as well as the random variable $k$ are positive, we can express the domain with $X^2+Y^2-K(\frac{\gamma\sigma^2_u}{p_t})<0$, which is equal to the domain $\mathbf{u}^T\mathbf{Q}_2\mathbf{u}+\mathbf{q_1}^T\mathbf{u}<0$, where $\mathbf{u}=[X, Y, K]^T$, $\mathbf{Q}_2=\begin{bmatrix}
1 & 0 & 0\\
0 &1 & 0\\
0 & 0& 0
\label{sigmad}
\end{bmatrix}$, and $\mathbf{q_1}=[0 , 0 ,-\frac{\gamma\sigma^2_u}{p_t}]^T$. Thus, the problem turns into finding the probability of $\mathbf{u}^T\mathbf{Q}_2\mathbf{u}+\mathbf{q_1}^T\mathbf{u}<0$ when $\mathbf{u}$ has the distribution $\mathcal{N}_3(N\mathbf{\mu_d},N\mathbf{\Sigma_d})$. We note that $\mathbf{u}=\mathbf{S}\mathbf{r}+N\mathbf{\mu_d}$, where $\mathbf{r}$ is a standard trivarate normal, and $\mathbf{S}=\sqrt{N}\mathbf{\Sigma_d}^{\frac{1}{2}} $. Thus, $\mathbf{r}=\mathbf{S}^{-1}(\mathbf{u}-N\mathbf{\mu_d})$, which cause a transformation to the integral domain as $\mathbf{r}^T\mathbf{\tilde{Q}}_2\mathbf{r}+\mathbf{\tilde{q_1}}^T\mathbf{r}<0$, where $\mathbf{\tilde{Q}}_2=\mathbf{S}\mathbf{Q}_2\mathbf{S}$,  and $\mathbf{\tilde{q_1}}=2\mathbf{S}\mathbf{Q}_2N\mathbf{\mu_d}+\mathbf{S}\mathbf{q}_1$.  Thus, the problem turns into finding the probability of standard normal $\mathbf{r}$ in the domain $\mathbf{r}^T\mathbf{\tilde{Q}}_2\mathbf{r}+\mathbf{\tilde{q_1}}^T\mathbf{r}<0$. Next, by eigenvalue decomposition of $\mathbf{\tilde{Q}}=\mathbf{V}\mathbf{\tilde{D}}\mathbf{V}^T$, in which $\mathbf{V}$ is orthogonal, we use another transformation, $\mathbf{t}=\mathbf{V}^T\mathbf{r}$, which is also standard trivariate normal. It results in a transformation in the integral domain as $\mathbf{t}^T\mathbf{\tilde{D}}\mathbf{t}+\mathbf{\tilde{a}}^T\mathbf{t}=\sum_{i}^{}\tilde{D}_i\chi'^{2}_{1,(\frac{\tilde{a}_i}{2\tilde{D}_i})^2}+\tilde{b}<0$, where $\mathbf{\tilde{a}}=\mathbf{V}^T\mathbf{\tilde{q_1}}$, $\tilde{b}\sim \mathcal{N}(m,s)$, $\chi'^{2}$ are chi-square variables with 1 degree of freedom, and $\tilde{D}_i$ are the diagonal element of $\mathbf{\tilde{D}}$. Thus, $\mathbf{t}^T\mathbf{\tilde{D}}\mathbf{t}+\mathbf{\tilde{a}}^T\mathbf{t}$, in which $\mathbf{t}$ is standard trivariate normal, is a generalized chi-square variable \footnote{One can calculate its parameters using the formula at \cite{Amethodtointegrate} or with the help of MATLAB toolbox provided by its authors.}and the problem turns into finding the cumulative density function (CDF) of this variable at zero.
\subsection{OP of the target}\label{opoftarget}
Based on (\ref{cramer}), (\ref{w}), and with the assumption of $\mathbf{b}(\theta)$ having the same configuration as $\mathbf{a}(\theta)$, we have:
\begin{align}
&\text{CRB}(\theta)=\frac{\sigma_r^2}{2 L p_t|\alpha|^2 || \mathbf{b}^. ||^2 | \mathbf{a}^H \mathbf{w} | ^2}\nonumber\\
&=\frac{6 \sigma_r^2}{L p_t|\alpha|^2 (M-1)(M)(M+1)\pi^2 \cos^2(\theta) }\frac{|b_1\mathbf{h}+b_2\mathbf{a}|^2}{|\mathbf{a}^H(b_1\mathbf{h}+b_2\mathbf{a})|^2}\nonumber\\
&\overset{(a)}{=}g(\theta)\frac{\sum_{i=1}^{N}|b_1h_i+b_2e^{-jf_i}|^2}{|\sum_{i=1}^{N}(b_1e^{jf_i}h_i)+b_2N|^2}\nonumber\\
&\overset{(b)}{=}\frac{Kg(\theta)}{\tilde{X}^2+\tilde{Y}^2+2N\mathcal{R}(b_2)\tilde{X}+2N\mathcal{I}(b_2)\tilde{Y}+N^2|b_2|^2} ,\label{cramer2}
\end{align}
where ($a$) is due to defining $g(\theta)\triangleq \frac{6 \sigma_r^2}{L p_t|\alpha|^2 (M-1)(M)(M+1)\pi^2 \cos^2(\theta)}$, ($b$) is due to defining $\tilde{X}=\sum_{i=1}^{N}\tilde{x}_i$, $\tilde{Y}=\sum_{i=1}^{N}\tilde{y}_i$ and $K=\sum_{i=1}^{N}k_i$, where $\tilde{x}_i=\mathcal{R}(b_1e^{jf_i}h_i)$, $\tilde{y}_i=\mathcal{I}(b_1e^{jf_i}h_i)$, and $k_i$ is defined in subsection \ref{opofuser}. Therefore, the OP of the target, $P_c$, is as (\ref{outagec}).
\begin{figure*}
\normalsize
\begin{align}
\mathbf{\Sigma_d}=\begin{bmatrix}
\frac{|b_2|^2}{2}+|b_1|^2\cos^2(\phi_1) & |b_1|^2\cos(\phi_1)\sin(\phi_1) & |b_1|(|b_1|^2+|b_2|^2)\cos(\phi_1)\\
|b_1|^2\cos(\phi_1)\sin(\phi_1) & \frac{|b_2|^2}{2}+|b_1|^2\sin^2(\phi_1) & |b_1|(|b_1|^2+|b_2|^2)\sin(\phi_1)\\
|b_1|(|b_1|^2+|b_2|^2)\cos(\phi_1) & |b_1|(|b_1|^2+|b_2|^2)\sin(\phi_1)& (|b_1|^4+2|b_1|^2|b_2|^2)
\label{sigmad}
\end{bmatrix}
\end{align}
\begin{align}
&P_c=P(\text{CRB}(\theta)>\epsilon)=\int_{0}^{\pi}P(\frac{Kg(\theta)}{\tilde{X}^2+\tilde{Y}^2+2N\mathcal{R}(b_2)\tilde{X}+2N\mathcal{I}(b_2)\tilde{Y}+N^2|b_2|^2}>\epsilon)|\theta)f_{\theta}(\theta)d\theta\label{outagec}
\end{align}
\hrulefill
\end{figure*}
In order to calculate the inner probability of (\ref{outagec}), we need to derive the joint PDF of $\tilde{X}$, $\tilde{Y}$, and $K$. We note that $\tilde{X}$, $\tilde{Y}$, and $K$ (also $\tilde{x}_i$, $\tilde{y}_i$, and $k_i$) are not independent, as they are functions of $h_i$ and $f_i$. By conditioning on $\theta$, $f_i$, $\forall i=1,...,N$, will be treated as constant in the following. We define $N$ random vectors, $\mathbf{\tilde{d}}_i=[\tilde{x}_i, \tilde{y}_i, k_i]^T \in \mathbb{R}^{3 \times 1}$. For any pair of $j$ and $i\neq j$ the vectors $\mathbf{\tilde{d}}_j$ and $\mathbf{\tilde{d}}_i$ are independent and have identical distribution because $h_i$s are i.i.d.. By using multidimensional CLT, when $N$ is large, we have:
$\sum_{i=1}^{N}\mathbf{\tilde{d}}_i\overset{(d)}{\rightarrow} \mathcal{N}_3(N\mathbf{\tilde{\mu}}_d,N\mathbf{\tilde{\Sigma}}_d)$,
where $\mathbf{\tilde{\mu}}_d$ and $\mathbf{\tilde{\Sigma}}_d$ are mean vector and covariance matrix of $\tilde{\mathbf{d}}_i$ (the same for $i=1,...,N$), respectively. Therefore, $[\tilde{X},\tilde{Y}, K]^T\overset{(d)}{\rightarrow} \mathcal{N}_3(N\mathbf{\tilde{\mu}}_d,N\mathbf{\tilde{\Sigma}}_d)$. Thus, by finding $\mathbf{\tilde{\mu}_d}$ and $\mathbf{\tilde{\Sigma}_d}$ with the aid of Lemma \ref{lemma2} (proof in Appendix \ref{lemma2p}), the joint PDF of $\tilde{X}$, $\tilde{Y}$, and $K$ is derived.
\begin{lemma}\label{lemma2}
$\mathbf{\tilde{\mu}}_d$ and $\mathbf{\tilde{\Sigma}}_d$ are calculated as (\ref{mudt}) and (\ref{sigmadt}), respectively:
\begin{align}
&\mathbf{\tilde{\mu}}_d=[0 ,0 , |b_1|^2+|b_2|^2]^T\label{mudt}
\end{align}
\begin{align}
\mathbf{\tilde{\Sigma}}_d\!=\!\!\!\!\begin{bmatrix}
\frac{|b_1|^2}{2} & 0 & |b_1|^2|b_2|\cos(\phi_2)\\
0 & \frac{|b_1|^2}{2} & |b_1|^2|b_2|\sin(\phi_2)\\
|b_1|^2|b_2|\cos(\phi_2) & |b_1|^2|b_2|\sin(\phi_2)& (|b_1|^4+2|b_1|^2|b_2|^2)
\label{sigmadt}
\end{bmatrix}
\end{align}
\end{lemma}
We note that as derived in Lemma \ref{lemma2}, the joint conditional PDF of $\tilde{X}$, $\tilde{Y}$, and $K$ is independent of $\theta$ and $\phi_1$. Using this and uniform distribution for $\theta$, and defining the domain $\mathcal{D}(\theta, \tilde{X},\tilde{Y}, K)= \frac{K}{\tilde{X}^2+\tilde{Y}^2+2N\mathcal{R}(b_2)\tilde{X}+2N\mathcal{I}(b_2)\tilde{Y}+N^2|b_2|^2}>\frac{\epsilon}{g(\theta)}$, (\ref{outagec}) will be:
\begin{align}
P_{c}&=\frac{1}{\pi}\!\!\int_{0}^{\pi}\!\!\iiint_{\mathcal{D}(\theta, \tilde{X},\tilde{Y}, K)}f(\tilde{X},\tilde{Y},K)\,d\tilde{X}\,d\tilde{Y}\,dK,d\theta,\label{outagec2}
\end{align}
where $f(\tilde{X},\tilde{Y},K)$ is the PDF of a trivariate normal distribution with a mean vector of $N\mathbf{\tilde{\mu}_d}$ and a covariance matrix of $N\mathbf{\tilde{\Sigma}_d}$ \footnote{We remark that by deriving $f(\tilde{X},\tilde{Y},K)$, we can also calculate the ergodic CRB as $E[\log^{1+\text{CRB}}]=\frac{1}{\pi}\int_{0}^{\pi}\int_{0}^{\infty}\!\!\iiint_{\mathcal{\tilde{D}}(t,\theta, \tilde{X},\tilde{Y}, K)}f(\tilde{X},\tilde{Y},K)\,d\tilde{X}\,d\tilde{Y}\,dKdtd\theta$ where $\mathcal{\tilde{D}}(t, \theta, \tilde{X},\tilde{Y}, K)=\frac{K}{\tilde{X}^2+\tilde{Y}^2+2N\mathcal{R}(b_2)\tilde{X}+2N\mathcal{I}(b_2)\tilde{Y}+N^2|b_2|^2}>\frac{\tilde{(2^t-1)}}{g(\theta)}$}. For each $\theta$, we can follow the method of Subsection \ref{opofuser} and write the domain $\mathcal{D}(\theta, \tilde{X},\tilde{Y}, K)$ in a quadratic form. Then, by calculating the parameters of the generalized chi-square variable obtained from some transformation on the domain, the inner triple integral will be the complement of the CDF (a function of $\theta, \tilde{X},\tilde{Y}, K$) of this generalized chi-square variable at zero. Finally, the outer integral on $\theta$ is calculated.
\section{Numerical Results}\label{simulations}
\begin{figure*}
\begin{multicols}{4}
    \includegraphics[scale=.25]{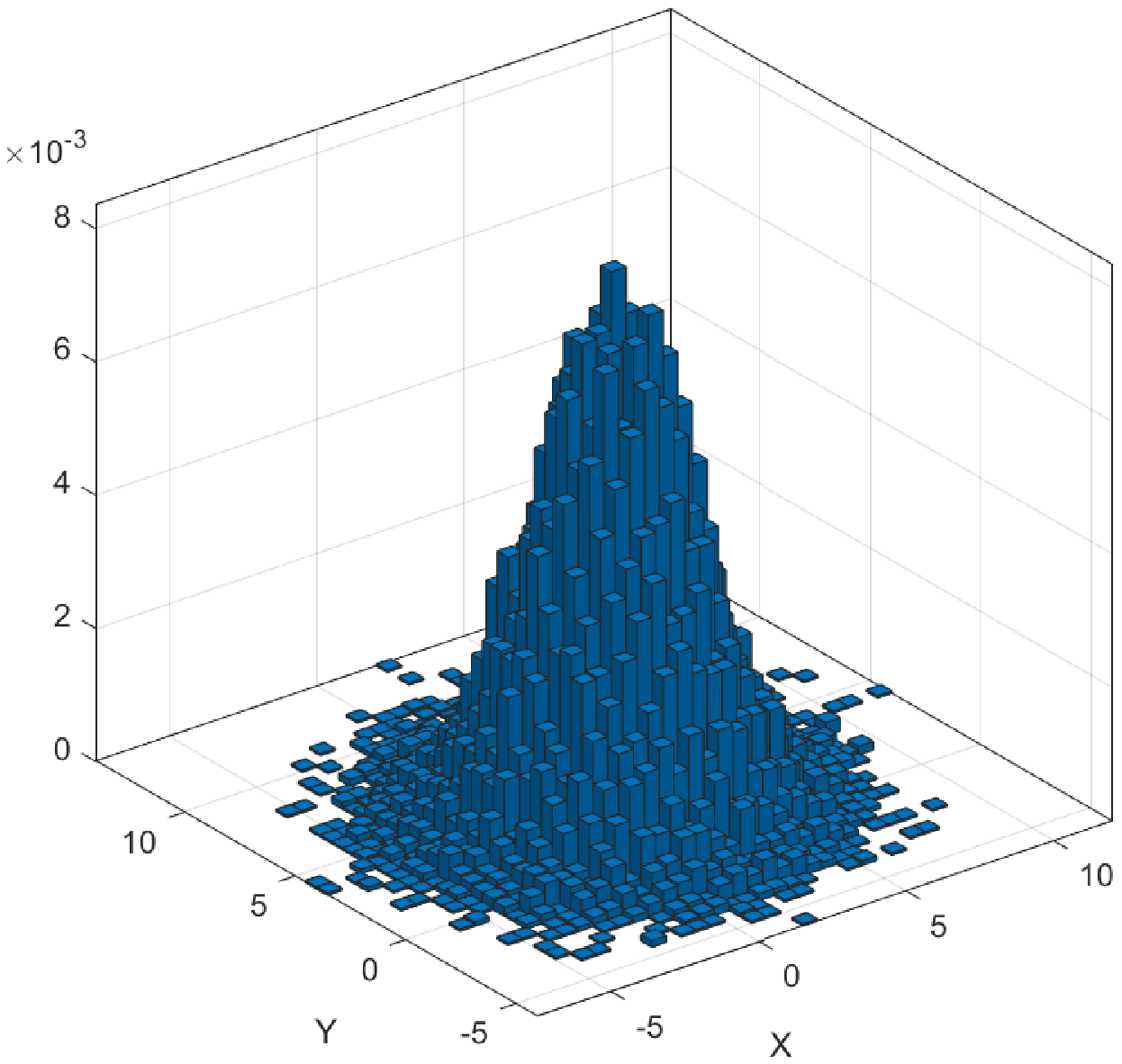}\par 
    \caption{Histogram of the joint PDF of $X$ and $Y$ with parameter $|b_1|=.2$, $|b_2|=.8$, $\phi_1=\frac{\pi}{3}$, and $\phi_2=0$.}\label{0}
    \includegraphics[scale=.25]{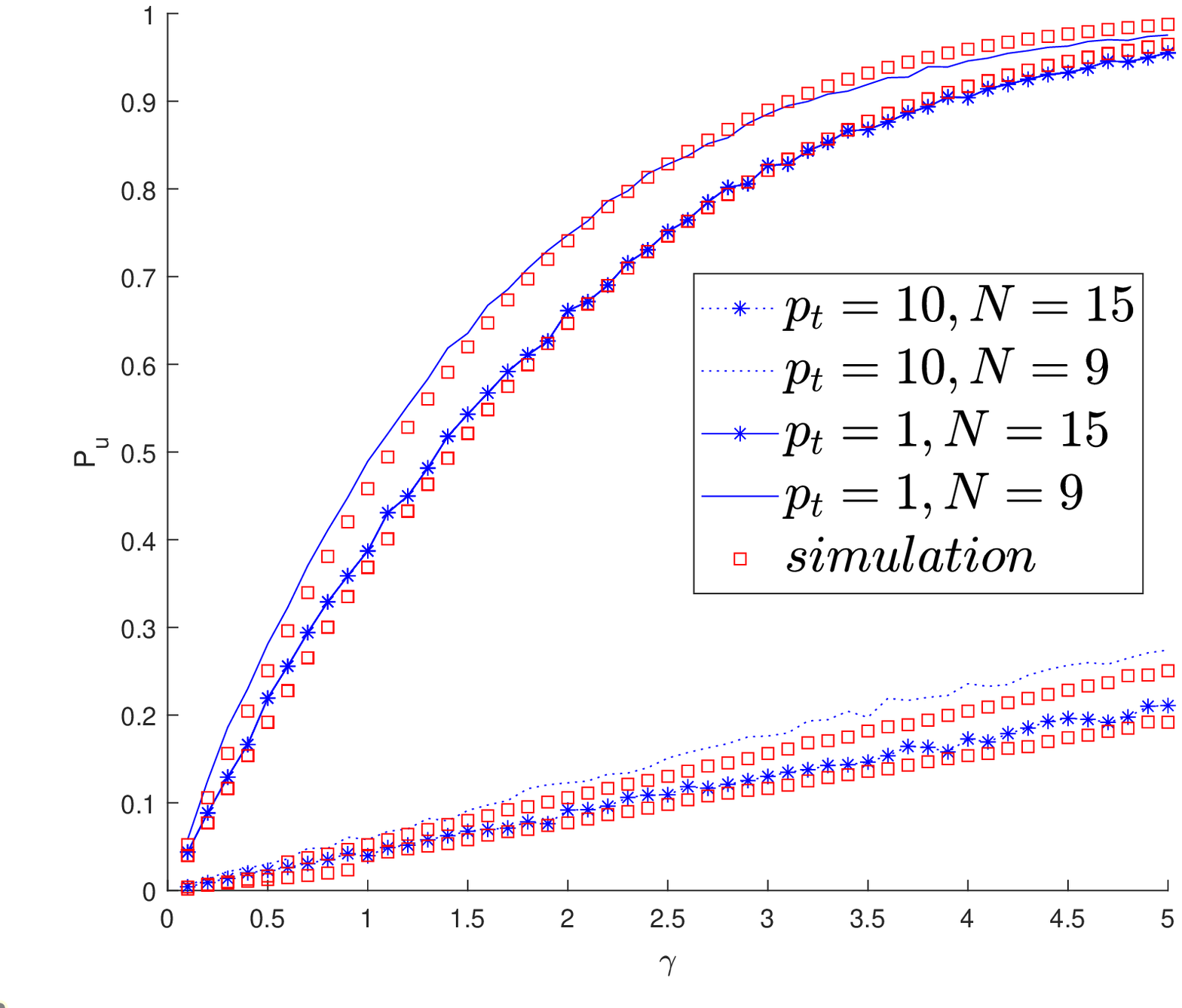}\par
    \caption{OP of the user versus $\gamma$ for different $N$ and $p_t$.}\label{1}
    \includegraphics[scale=.25]{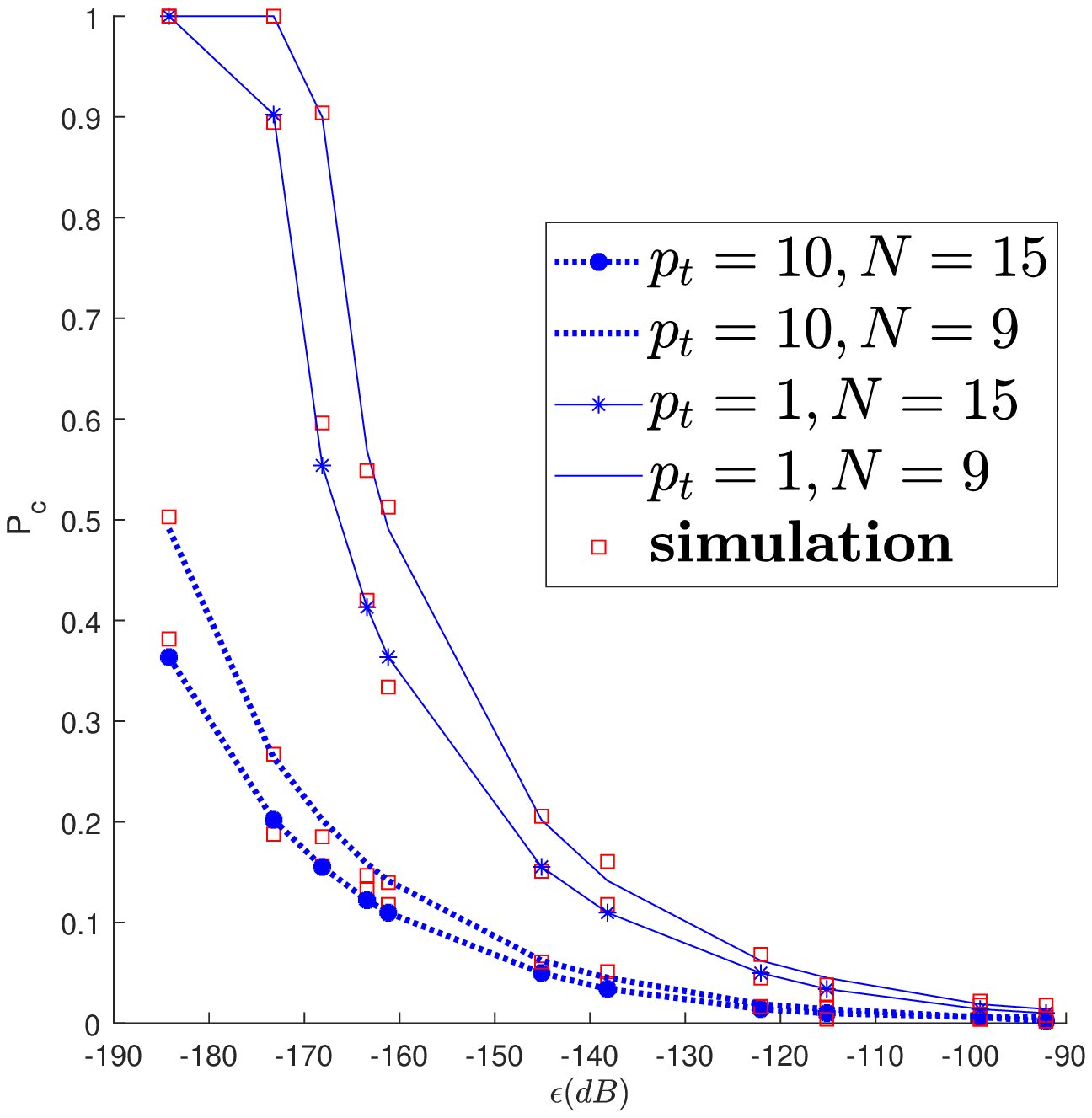}\par 
    \caption{OP of the target versus $\epsilon(dB)$ for different $N$ and $p_t$.}\label{2}
    \includegraphics[scale=.25]{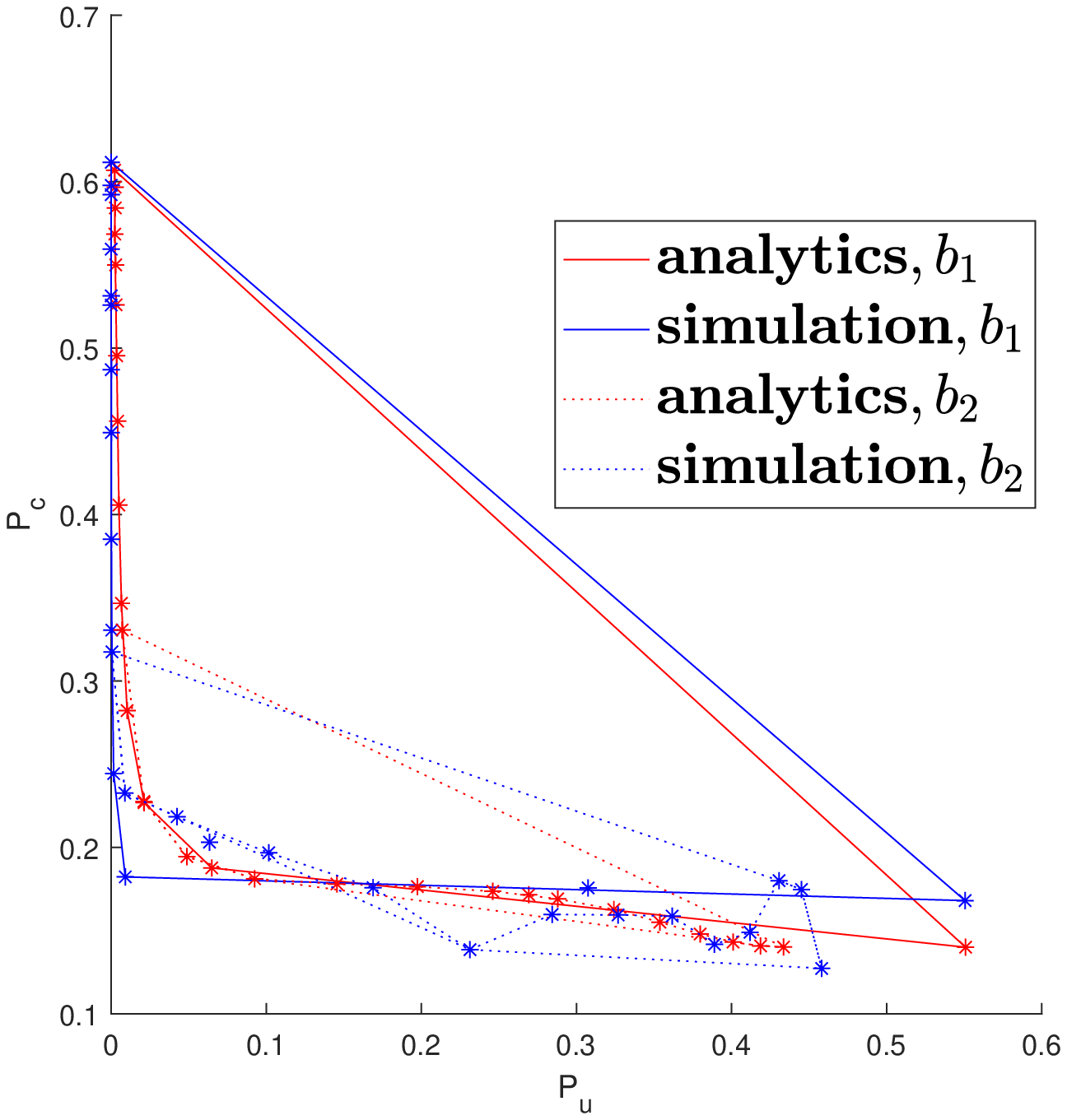}\par
    \caption{OP of target versus OP of the user for different $|b_1|$ and and $|b_2|$}\label{3}
    \end{multicols}
\end{figure*}

We consider a BS with
$N=15$ (unless stated otherwise) and $M=17$ transmitter and receiver antennas.
The power budget and the noise variance at the user and BS are $p_t=10$ (unless stated otherwise), $\sigma_r=1$, and $\sigma_u=1$, receptively. The length of the radar frame and the reflection coefficient are $L = 30$ and $\alpha=1$, respectively. the beamforming vector parameters are $|b_1|=.2$, $|b_2|=.8$, $\phi_1=\frac{\pi}{3}$, and $\phi_2=0$ (unless sated otherwise). The simulation results are based on $10000$ randomly seeded channel realizations.
It is not possible to plot the joint PDF of $X$, $Y$, and $Z$, defined in (\ref{opofuser}). However, Fig. \ref{0}, created by Mont-Carlo simulation, shows the joint PDF of $X$ and $Y$, which confirms the numerical approximation \footnote{The joint PDF of $X$ and $K$, as well as $Y$ and $K$, is similar to Fig. \ref{0}, with different mean vector and covariance matrix. Due to the space limitations, we ignored it.}.

Fig. \ref{1} and Fig. \ref{2} show the OP of the user, $P_u$, and OP of the target, $P_c$, respectively, versus $\gamma$ and $\epsilon$ (the thresholds of the OP) for two different value of $N$ and $p_t$. As expected, with increasing $\gamma$ ($\epsilon$), the OP of the user (target) increases (decreases). Moreover, by increasing $p_t$, both the OP of the user and the target decrease due to the increment in the received signal by user and reflected signal by the target. Furthermore, as the number of antennas $N$ increases, both the OP of the user and the target decrease due to a more focused and directed beamforming vector. It is worth noting that even when by reducing the number of antennas to $N=9$, the simulation results closely match the analytical results, indicating that the CLT still holds even for a small number of antennas. In summery, increasing $N$ and $p_t$ improve the system performance in both communication and sensing aspects.

To demonstrate the trade off between S \& C, Fig. \ref{3}, shows the achievable region of the OP of the target and the OP of the user based on $|b_1|$ (solid lines) and $|b_2|$ (dashed lines) when $N=9$. The OP of the target and the user are calculated for different value of $|b_1|$ and $|b_2|$. Our parameters are as before, except for $\gamma=8$ and $\epsilon=8\times 10^{-7}$. This figure reveals that by increasing $|b_1|$ (decreasing $|b_2|$) (when moving to the left-hand side of the figure, $|b_1|$ increases and $|b_2|$ decrease), the Op of the user (target) decreases (increases), showing an improvement in the communication performance and a degradation in sensing performance. This is because by increasing $|b_1|$ (decreasing $|b_2|$), we steer the beamforming vector toward user, resulting in a higher error estimation for $\theta$.
\section{Conclusion}\label{conclusion}
We analysed the fundamental performance of a downlink MIMO ISAC system and investigated the impact of randomness in the channels. The OP of the user for communication and the OP of the target were derived as our performance metrics. We used a MRT beamformer to share resources between S \& C and observed the trade-off between them. By allocating
more power towards the user direction, the OP of the user decreased, indicating an improvement in communication performance. However, the OP of the target increased, implying a higher error in angle estimation for the target.
\bibliographystyle{ieeetr}
\bibliography{ref}

\appendices

\section{Proof of Lemma \ref{lemma1}}\label{lemma1p}
First, we expand the variable $b_1|h_i|^2+b_2h^*_ie^{-jf_i}=x_i+jy_i$, where $h_i=m_i+jn_i$ is defined at section \ref{systemmodel}, and $x_i$, $y_i$, and $k_i$ are defined at section \ref{opofuser}. (for simplification, we omit the subscript $i$ as $\mathbf{d}_i=[x_i,y_i,k_i]^T$s for $i=1,...,N$ are i.i.d. and have the same mean vector and covariance matrix):
\begin{align}
\!\!\!\!b_1|h|^2+b_2h^*e^{-jf}\!\!&=|b_1|e^{j\phi_1}|h|^2\!\!\!+|b_2|e^{j(\phi_2-f)}(m-jn)
\end{align}
After some simple calculation and using $\cos(a+b)=\cos(a)\cos(b)-\sin(a)\sin(b)$ and $\sin(a+b)=\sin(a)\cos(b)+\cos(a)\sin(b)$, the real and imaginary parts of this variable is derived as (\ref{xy}). Moreover, $k$ is shown at (\ref{k}).
\begin{figure*}[!t]
\normalsize
\begin{align}
x&=\mathcal{R}(b_1|h|^2+b_2h^*e^{-jf})=|b_1|(m^2+n^2)\cos(\phi_1)+|b_2|m\cos(\phi_2-f)+|b_2|n\sin(\phi_2-f),\nonumber\\
y&=\mathcal{I}(b_1|h|^2+b_2h^*e^{-jf})=|b_1|(m^2+n^2)\sin(\phi_1)+|b_2|m\sin(\phi_2-f)-|b_2|n\cos(\phi_2-f),\label{xy}\\
k&=|b_1h+b_2e^{-jf}|^2=m^2|b_1|^2+n^2|b_1|^2+2m|b_1b_2|\cos(-\phi_1+\phi_2-f)+2n|b_1b_2|\sin(-\phi_1+\phi_2-f)+|b_2|
^2.\label{k}
\end{align}
\hrulefill
\end{figure*}
Next, we calculate the mean vector $\mu_d=[E\{x\}, E\{y\},E\{k\}]^T$. The moments of a normal random variable, $e \!\sim \mathcal{N}(0,\sigma^2\!)$, are \cite{moment}:
\begin{align}
E\{e^{2k-1}\}=0 ,\quad E\{e^{2k}\}=\sigma^{2k}\frac{(2k)!}{2^k k!}.
\label{z^k}
\end{align}
By using (\ref{z^k}), and the fact that $m$ and $n$ are i.i.d with the distribution of $\mathcal{N}(0,\frac{1}{2})$, we have:
\begin{align}
E\{m\}&=E\{n\}=E\{m^3\}=E\{n^3\}=0;\nonumber\\ E\{m^2\}&=E\{n^2\}=\frac{1}{2}; \quad E\{m^4\}=E\{n^4\}=\frac{3}{4}.\label{moment}
\end{align}
Since expectation is a linear operation and by using (\ref{moment}) and the fact that all other variables except $m$ and $n$ in (\ref{xy}) and (\ref{k}) are constant (conditioned on $\theta$), we have:
\begin{align}
E\{x\}=|b_1|\cos(\phi_1); E\{y\}=|b_1|\sin(\phi_1);E\{k\}=|b_1|^2+|b_2|^2;\label{mudp}
\end{align}
Next, we calculate the covariance matrix $\mathbf{\Sigma_d}$. By definition, we have:
\begin{align}
\mathbf{\Sigma_d}=\begin{bmatrix}
\text{var}(x) & \text{cov}(xy) & \text{cov}(xk)\\
\text{cov}(xy) & \text{var}(y) & \text{cov}(yk)\\
\text{cov}(xk) & \text{cov}(yk) &\text{var}(k)
\end{bmatrix}
\end{align}
where \text{var(.)} and \text{cov(.)} denotes variance and covariance. Using (\ref{moment}), (\ref{mudp}), the definition of covariance, and after some mathematical derivations, we can calculate each element of the covariance matrix as in (\ref{sigmad}). The proof is complete.
\section{Proof of Lemma \ref{lemma2}}\label{lemma2p}
First, we have $b_1h_ie^{jf_i}=|b_1|e^{j(\phi_1+f_i)}(m_i+jn_i)=\tilde{x}_i+j\tilde{y}_i$, where $\tilde{x}_i$, $\tilde{y}_i$, are defined at section \ref{opoftarget} (for brevity, we omit the subscript $i$ as $\mathbf{\tilde{d}}_i=[\tilde{x}_i,\tilde{y}_i,k_i]^T$s for $i=1,...,N$  are i.i.d with the same mean vector and covariance matrix). After some simple calculations, we have:
\begin{align}
\tilde{x}&=|b_1|\cos(\phi_1+f)m-|b_1|n\sin(\phi_1+f),\nonumber\\
\tilde{y}&=|b_1|\cos(\phi_1+f)n+|b_1|m\sin(\phi_1+f).\label{xyt}
\end{align}
Next, we calculate the mean vector $\mathbf{\tilde{\mu}}_d=[E\{\tilde{x}\},E\{\tilde{y}\},E\{k\}]^T$.
Since expectation is a linear operation and by using (\ref{moment}) and the fact that all other variables except $m$ and $n$ in (\ref{xyt}) and (\ref{k}) are constant (conditioned on $\theta$), we have:
\begin{align}
E\{\tilde{x}\}=0; E\{\tilde{y}\}=0.\label{mudpt}
\end{align}
Using (\ref{moment}), (\ref{mudp}), (\ref{mudpt}), and the definition of covariance, and after some mathematical derivations, we can calculate each element of the covariance matrix $\mathbf{\tilde{\Sigma}_d}$ as in (\ref{sigmadt}). The proof is complete.

\end{document}